\documentstyle[prl,aps,multicol,graphicx,amssymb]{revtex}
\newcommand{\bleq}{\ifpreprintsty \else
\end{multicols}\vspace*{-3.5ex}{\tiny \noindent\begin{tabular}[t]{c|}
\parbox{0.493\hsize}{~} \\ \hline \end{tabular}} \fi}
\newcommand{\eleq}{\ifpreprintsty \else
{\tiny\hspace*{\fill}\begin{tabular}[t]{|c}\hline
\parbox{0.49\hsize}{~} \\
\end{tabular}}\vspace*{-2.5ex}\begin{multicols}{2} \fi}
\newcommand{\bcols}{\ifpreprintsty\else\begin{multicols}{2}\fi}
\newcommand{\ecols}{\ifpreprintsty\else\end{multicols}\fi}
\renewcommand{\vec}[1]{{\bf #1}}
\renewcommand{\mathbf}{\bf\mathsf} \let\oldhat=\hat
\renewcommand{\hat}[1]{\oldhat{\bf #1}}

\begin{document}
\draft 
\title{Lyapunov Exponent Pairing for a Thermostatted Hard-Sphere Gas
under Shear in the Thermodynamic Limit}
\author{Debabrata Panja$^{*}$ and Ramses van Zon$^{\dagger}$} 
\maketitle 
\begin{center}
{\em $^{*}$Instituut Lorentz, Universiteit Leiden, Postbus 9506, 2300
RA Leiden, The Netherlands\\ $^{\dagger}$ Chemical Physics Theory
Group, Chemistry Department, University of Toronto,\\80 St. George
St. Toronto, Ontario M5S 3H6, Canada\/}
\end{center}

\begin{abstract}
We demonstrate why for a sheared gas of hard spheres, described by the
SLLOD equations with an iso-kinetic Gaussian thermostat in between 
collisions, deviations of the conjugate pairing rule for the Lyapunov 
spectrum are to be expected, employing a previous result that for a 
large number of particles $N$, the iso-kinetic Gaussian thermostat is 
equivalent to a constant friction thermostat, up to $1/\sqrt{N}$ 
fluctuations. We also show that these deviations are at most of the 
order of the fourth power in the shear rate.
\end{abstract}


\bcols
\noindent 

The SLLOD equations of motion, combined with Lees-Edwards boundary
condition\cite{LE_jpa_72}, were originally proposed in
Refs. \cite{EM_cpr_84,EM_pra_84}, and since then they have been
convenient tools to calculate the shear viscosity of gases in the bulk
by means of non-equilibrium molecular dynamics simulations for many
years. These studies consider systems with a large number of mutually
interacting particles that are driven by an external shear rate
$\gamma$ \cite{Morriss_pra_88,Morriss_pla_89,EM_book_academic}. In
these studies, the iso-kinetic Gaussian thermostat is an artificial
way to continuously remove the energy generated inside the system due
to the work done on it by the external shear field, such that a
non-equilibrium steady state, homogeneous in space, can be
reached. The Lyapunov spectrum of such systems is of interest since it
has been shown that the shear viscosity can be related to the spectrum
\cite{Morriss_pla_89,EM_book_academic}, which can be numerically
obtained as a function of the shear rate\cite{ignorant}.  The analysis
of the simulation data \cite{Morriss_pla_89} indicated that the sum of
the largest and the smallest, the sum of the second largest and the
second smallest and so on, were the same. The phenomenon of such
pairing of the Lyapunov exponents is known as the Conjugate Pairing
Rule, or the CPR. Based on this observation, an attempt to prove an
exact CPR was made for arbitrary inter-particle potentials and
arbitrary $\gamma$ \cite{ECM_pra_90,SEM_pra_92}, and later studies and
better simulation techniques \cite{ISE_physica_97,SEI_chaos_98}
indicated that for systems obeying the SLLOD equations of motion, the
CPR is not satisfied exactly under these general
conditions\cite{morrisscomment}.  However, any conclusive {\it
theoretical}\, proof regarding the status of an approximate CPR for
systems under SLLOD equations of motion is absent in the literature
till now, leaving the problem open for a long time.

The SLLOD equations of motion describe the dynamics of a collection of
$N$ particles constituting a fluid with a macroscopic velocity field
$\vec u(\vec r)=\gamma y \hat x$.  For particles of unit mass, the
equations of motion of the $i$-th particle, in terms of its position
$\vec{r}_{i}$ and peculiar momentum $\vec{p}_{i}$, is given by
\begin{eqnarray}
\dot{\vec{r}}_{i}\,=\,\vec{p}_{i}\,+\,\gamma
y_{i}\hat{x}\,,\hspace{0.6cm}
\dot{\vec{p}}_{i}\,=\,\vec{F}_{i}\,-\,\gamma
p_{iy}\hat{x}\,-\,\alpha\vec{p}_{i}\,,
\label{e1}
\end{eqnarray}
where $\vec{F}_{i}$ is the force on the $i$-th particle due to the
other particles in the system. The value of $\alpha$, the coefficient
of friction representing the iso-kinetic Gaussian thermostat, is
chosen such that the total peculiar kinetic energy of the system,
$\sum_{i}{p}^{2}_{i}/2$, is a constant of motion in between collisions. 
In terms of the positions $\vec{r}_{i}$ and the laboratory momenta 
$\vec{v}_{i}$ of the particles, Eq.~(\ref{e1}) reads
\begin{eqnarray}
\dot{\vec{r}}_{i}\,=\,\vec{v}_{i}\,,\hspace{0.6cm}
\dot{\vec{v}}_{i}\,=\,\vec{F}_{i}\,+\,
\alpha
\gamma
y_{i}\hat{x}\,-\,
\alpha
\vec{v}_{i}\,.
\label{e2}
\end{eqnarray}

In the present context, the gas particles are hard spheres, which for
simplicity are assumed to have unit radius.  The dynamics of the gas
particles consists of an alternating sequence of flight segments and
instantaneous {\it binary\/} collisions. During a flight, the dynamics
of the gas particles is described by Eqs.~(\ref{e2}) with
$\vec{F}_{i}=0$. 
At an instantaneous collision between the $i$-th and
the $j$-th sphere, the post-collisional positions and laboratory
momenta ($+$ subscripts) are related to their pre-collisional values
($-$ subscripts) by
\begin{eqnarray}
\vec{r}_{i+}&=&\vec{r}_{i-},\quad\quad\vec{r}_{j+}\,=\,\vec{r}_{j-}\,,\nonumber\\
\vec{v}_{i+}&=&\vec{v}_{i-}\,-\,\{(\vec{v}_{i-}\,-\,\vec{v}_{j-})\cdot\hat
{n}_{ij}\}\,\hat{n}_{ij}\quad\mbox{and}\nonumber\\
\vec{v}_{j+}&=&\vec{v}_{j-}\,+\,\{(\vec{v}_{i-}\,-\,\vec{v}_{j-})\cdot\hat
{n}_{ij}\}\,\hat{n}_{ij}\,,
\label{e5}
\end{eqnarray}
while the positions and the velocities of the rest of the spheres
remain unchanged. Here, $\hat{n}_{ij}$ is the unit vector along the
line joining the center of the $i$-th sphere to the $j$-th sphere at
the instant of collision. Note that because we applied the iso-kinetic
Gaussian thermostat only between collisions\cite{Petravic}, the
peculiar kinetic energy changes in individual collisions. These 
changes are random, both in magnitude and sign, due to the randomness 
of the collision parameters, and hence it is quite likely
that the system would reach a steady state, where the average change
of peculiar kinetic energy would be zero.

In terms of the $3N$-dimensional vectors
$\vec{R}=(\vec{r}_{1},\vec{r}_{2},...,\vec{r}_{N})$,
$\vec{V}=(\vec{v}_{1},\vec{v}_{2},...,\vec{v}_{N})$ and
$\hat{N}_{ij}$, whose $l$-th entry is given by $\hat{N}^{l}_{ij} =
(\delta_{l,i}-\delta_{l,j}) \,\hat{n}_{ij} / \sqrt{2}$ $(l=1,2,..,N)$,
Eqs.~(\ref{e2}-\ref{e5}) can be compacted to
\begin{eqnarray}
\dot{\vec{R}}\,=\,\vec{V}\,,\hspace{0.6cm}\dot{\vec{V}}\,=\,\,
\alpha
\gamma
{\mathbf C} \vec{R}\,-\,
\alpha
\vec{V}\,
\label{e6}
\end{eqnarray}
during a flight segment and
\begin{eqnarray*}
\vec{R}_{+}\,=\,\vec{R}_{-}\,,\hspace{0.6cm}\vec{V}_{+}\,=\,\vec{V}_{-}\,-\,2\,(\vec{V}_{-}\cdot\hat{N}_{ij})\,\hat{N}_{ij}\,
\end{eqnarray*}
at a collision between the $i$-th and the $j$-th
sphere\cite{Panja_preprint}. Here, ${\mathbf C}$ is a $3N\!\times\!3N$
matrix with $N\!\times\!N$ entries, each of which is a $3\!\times\!3$
matrix. In terms of the entry index $(l,m)$, in the $xyz$-basis,
${\mathbf C}_{lm}={\mathbf c}\,\delta_{lm}$ ($l,m=1,2,..,N$) and
\begin{eqnarray*}
{\mathbf c}&=&\hat x\hat
y\,=\,\left[\begin{array}{ccc}{0}\,\,\,\,{1}\,\,\,\,{0}\\{0}\,\,\,\,{0}\,\,\,\,{0}\\{0}\,\,\,\,{0}\,\,\,\,{0}\end{array}\right]\,.
\end{eqnarray*}

Having described the dynamics of the infinitesimal deviation
$\vec{\delta X}=(\vec{\delta R},\,\vec{\delta V})$ between two typical
trajectories in the $6N$-dimensional phase space for a time $t$ as
\begin{eqnarray}
\vec{\delta X}(t)&=&{\mathbf L}(t)\,\vec{\delta X}(0)\,,
\label{enew10}
\end{eqnarray}
the Lyapunov exponents for this system are the logarithms of the
eigenvalues of the matrix ${\mathbf \Lambda}$, defined by
\begin{eqnarray*}
{\mathbf \Lambda}\,=\,\lim_{t\rightarrow\infty}\left[{\mathbf
\tilde{L}}(t)\right] ^{1/(2\,t)},
\end{eqnarray*}
where ${\mathbf \tilde{L}}(t)=[{\mathbf{L}}(t)]^{\mbox{\scriptsize
T}}\,{\mathbf{L}}(t)$.

It can be shown\cite{PZ_long} that the  {\it sufficient condition\/}
for the CPR to hold exactly for a dynamical system obeying
Eq.~(\ref{enew10}) is the existence of a {\it constant non-singular}
matrix ${\mathbf K}$ satisfying ${\mathbf K}^{2}\!\propto\!{\mathbf
I}$, such that
\begin{eqnarray}
[{\mathbf L}(t)]^{\mbox{\scriptsize T}}\,{\mathbf K}\,{\mathbf
L}(t)=\mu{\mathbf K}.
\label{econd}
\end{eqnarray}
Here, $\mu$ is a scalar function of $t$. If ${\mathbf{L}}(t)$
satisfies Eq.~(\ref{econd}), then we call ${\mathbf{L}}(t)$ to be
``generalized $\mu$-symplectic''. It is easy to show from
Eq.~(\ref{econd}) that if $\tilde{L}$ is an eigenvalue of
${\mathbf\tilde{L}}(t)$, then so is $\mu^{2}/\tilde{L}$; from which
the (exact) CPR follows. For the situations where the CPR has been
proved to be exact
\cite{Panja_preprint,DM_pre_96,WL_cmp_98,Ruelle_jsp_99}, only the
$\mu$-symplecticity case of Eq.~(\ref{econd}) (i.e., ${\mathbf
K}={\mathbf J}$, where ${\mathbf J}$ is the usual symplectic matrix)
has been exploited. In this context, we note that despite the
similarity between the present problem and the one discussed in
Ref. \cite{Panja_preprint}, the elaborate formalism developed therein
is not applicable here.

A significant simplification can be achieved by noticing that the
coefficient of friction $\alpha$, in the non-equilibrium steady state,
fluctuates with $1/\sqrt{N}$ fluctuations around a fixed value
$\alpha_{0}$ in the thermodynamic limit \cite{Zon_pre_99}. Thus, to
calculate the Lyapunov exponents for large $N$, to which we confine
ourselves henceforth, $\alpha$ can be replaced by $\alpha_{0}$ in
Eq.~(\ref{e6}), except for a beginning transient time.  On average,
for small $\gamma$, $\alpha\propto\gamma^2$ and so is
$\alpha_0$. Higher order corrections play a role for larger shear
rates.

In the following analysis, we first explore the status of the CPR when
the coefficient of friction is a constant, $\alpha_{0}$, and then
return to the case where the coefficient of friction represents an
iso-kinetic Gaussian thermostat. The detailed derivation of the
following results is given elsewhere\cite{PZ_long}. At present, we
focus only on the main points.

Once $\alpha_{0}$ replaces $\alpha$ in Eq.~(\ref{e6}), we find that in
the time evolution of $\vec{\delta X}$ over a collision-less  flight
segment between $t$ and $t+\tau$ is given by
\begin{eqnarray}
\vec{\delta X}(t+\tau)\,=\,{\mathbf H}(\tau)\,\vec{\delta X}(t)\,.
\label{e9}
\end{eqnarray}
${\mathbf H}(\tau)$ can be decomposed into $3N\times3N$ sub-matrices as
\begin{eqnarray}
{\mathbf H}(\tau) = \left[\begin{array}{cc}{\mathbf
h}^{[1]}(\tau)&{\mathbf h}^{[2]}(\tau)\\{\mathbf
h}^{[3]}(\tau)&{\mathbf h}^{[4]}(\tau)\\\end{array}\right]\,.
\label{Hform}
\end{eqnarray}
Having further decomposed each of the ${\mathbf h}^{[k]}(\tau)$
matrices ($k=1,\ldots,4$) into $N\times N$ entries of $3\times3$
matrices as ${\mathbf h}^{[k]}_{lm}(\tau)$ ($l$ and $m$ are counted
along the row and the column respectively), we have (with ${\mathbf
I}$ as the identity matrix)
\begin{eqnarray}
   {\mathbf h}^{[1]}_{lm}(\tau)&=&\left\{{\mathbf I} +
\left[\,\gamma\tau -\frac{\gamma}{\alpha_0}\,(1-e^{-\alpha_{0}\tau})
\right]{\mathbf c}\right\}\,\delta_{lm}\,, \nonumber \\ {\mathbf
h}^{[2]}_{lm}(\tau)&=&\bigg\{\frac{1-e^{-\,\alpha_{0}\tau}}{\alpha_{0}}
{\mathbf I} \nonumber\\&& + \frac{\gamma}{\alpha_0^2}\left[
\alpha_0\tau (1+e^{-\alpha_0\tau})-2+2e^{-\alpha_0\tau} \right]
{\mathbf c}\bigg\}\,\delta_{lm}\,, \nonumber \\  {\mathbf
h}^{[3]}_{lm}(\tau)&=&
\left\{\gamma\,[1-e^{-\alpha_{0}\tau}]\,{\mathbf
c}\right\}\,\delta_{lm} \quad\mbox{and}\nonumber \\  {\mathbf
h}^{[4]}_{lm}(\tau)&=& \left\{e^{-\alpha_{0}\tau} {\mathbf I}
-\gamma\left[\tau+{\frac{1} {\alpha_{0}}(1-e^{\alpha_{0}\tau})}\right]
{\mathbf c}\right\}\,\delta_{lm}\,,
\label{e4.3}
\label{splitg}
\end{eqnarray}
If we now form a $6N\times6N$ matrix ${\mathbf G}$, [in the notation
of Eq. (\ref{Hform})] which looks like ${\mathbf
G}^{[1]}_{lm}={\mathbf G}^{[4]}_{lm}=\O$ and ${\mathbf
G}^{[2]}_{lm}={\mathbf G}^{[3]}_{lm}={\mathbf g}\,\delta_{lm}$, where
\begin{eqnarray}
   {\mathbf g}= \left[\begin{array}{ccc} {0}&{1}&{0}\\ {1}&{0}&{0}\\
	{0}&{0}&{1} \end{array}\right]\,.
\label{e11}
\end{eqnarray}
Then ${\mathbf H}(\tau)$ can be easily shown to satisfy \cite{otherG}
\begin{eqnarray}
[{\mathbf H}(\tau)]^{\mbox{\scriptsize T}}\,{\mathbf G}\,{\mathbf
H}(\tau)\,=\,e^{-\alpha_{0}\tau}\,{\mathbf G}\,.
\label{e10}
\end{eqnarray}
Thus, ${\mathbf H}(\tau)$ is generalized $\mu$-symplectic with
${\mathbf G}$, but it is {\bf not} $\mu$-symplectic, i.e., $[{\mathbf
H}(\tau)]^{\mbox{\scriptsize T}}{\mathbf J}\,{\mathbf H}(\tau)\neq
e^{-\alpha_{0}\tau}{\mathbf J}$.  The fact that the same analysis
[Eqs.~(\ref{e9}-\ref{e11})] can be carried out for any constant
coefficient of friction (not necessarily $\alpha_{0}$), implies that
the CPR is exact for a {\it collision-less}  gas of point particles
obeying Eq.~(\ref{e6}) with a constant coefficient of friction.  This
has been found previously in simulation data \cite{SEI_chaos_98}.

For the transformation of $\vec{\delta X}$ over a binary collision
between the $i$-th and the $j$-th sphere, we follow the explicit
derivation in Ref. \cite{Panja_preprint}, which in turn is based on
the formalism developed simultaneously by Gaspard and
Dorfman\cite{GaspardDorfman}, and by Dellago and co-workers
\cite{DPH_pre_96}.  The post-collisional infinitesimal deviation
vector $\vec{\delta X}_{+}$ can be related to its pre-collisional
value $\vec{\delta X}_{-}$ by
\begin{eqnarray*}
\vec{\delta X}_{+}\,=\,{\mathbf M}_{ij}\,\vec{\delta X}_{-}\,,
\end{eqnarray*}
where the $6N\!\times\!6N$ matrix ${\mathbf M}_{ij}$ can be decomposed
into four $3N\!\times\!3N$ blocks, having the following structure
\begin{eqnarray*}
  {\mathbf M}_{ij}&=& ({\mathbf I}\,-\,2\hat{N}_{ij}\hat{N}_{ij})
	\left[\begin{array}{ccc}  {\mathbf I} && 0\\ {\mathbf
	R}&&{\mathbf I}
\end{array}
\right]\,.
\end{eqnarray*}
Here, ${\mathbf R}$ is a symmetric matrix. The above expression for
${\mathbf M}$ implies that the collisions are symplectic, i.e.,
\begin{eqnarray*}
{\mathbf M}_{ij}^{\mbox{\scriptsize T}}{\mathbf J}\,{\mathbf
M}_{ij}={\mathbf J}\,,
\end{eqnarray*}
but {\bf not} generalized symplectic with ${\mathbf G}$ (${\mathbf
M}^{\mbox{\scriptsize T}}_{ij}{\mathbf G}{\mathbf M}_{ij}\neq{\mathbf
G}$).

We can now express the matrix ${\mathbf L}(t)$ in terms of the
${\mathbf H}$ and ${\mathbf M}$ matrices in the following way: if the
dynamics involves free flight segments separated by $s$ instantaneous
binary collisions at $t_{1}$, $t_{2},\ldots,t_{s}$ such that
$0<t_{1}<t_{2}<\ldots<t_s<t$, then
\begin{eqnarray}
   {\mathbf L}(t)= {\mathbf H}(\Delta t_{s})\,{\mathbf M}_{i_sj_s}\,
	{\mathbf H}(\Delta t_{s-1})  \cdots {\mathbf
	M}_{i_1j_1}\,{\mathbf H}(\Delta t_{0})\,.
\label{e15}
\end{eqnarray}
Here, $\Delta t_s=t-t_s$ and $\Delta t_i=t_{i+1}-t_i$ for
$i=1,\ldots,(s-1)$.

The consequences of Eqs. (\ref{e9}-\ref{e15}) can be summarized by the
following: for a collection of hard spheres obeying the SLLOD
equations of motion with constant coefficient of friction
$\alpha_{0}$, {\it (a)\/} the ${\mathbf H}$ matrices are generalized
$\mu$-symplectic with ${\mathbf G}$, but {\bf not} with ${\mathbf J}$,
and {\it (b)\/} the ${\mathbf M}$ matrices are symplectic but {\bf
not} generalized $\mu$-symplectic with ${\mathbf G}$.  Hence, once the
${\mathbf H}$ and the ${\mathbf M}$ matrices are combined together, as
in Eq.~(\ref{e15}), ${\mathbf L}(t)$ is seen to be generalized
$\mu$-symplectic with neither ${\mathbf G}$ nor ${\mathbf J}$. This is
consistent with the claim that ${\mathbf L}(t)$ is {\bf not}
generalized $\mu$-symplectic (and consequently, the CPR does not hold
exactly) for a collection of hard spheres obeying the SLLOD equations
of motion with constant coefficient of friction $\alpha_{0}$.

The degree of deviation from an exact CPR must follow from the
properties of ${\mathbf L}(t)$, and to estimate this deviation, we can
use either ${\mathbf K}={\mathbf G}$, or ${\mathbf K}={\mathbf J}$ in
Eq.~(\ref{econd}). While the former choice implies that one has to try
to estimate the deviation from an exact CPR from the distribution of
the unit vectors $\hat{N}_{ij}$'s and the collision angles for
different sets of binary collisions in the expression of ${\mathbf
M}_{ij}$'s, the latter choice means that one can make the estimate by
using the typical magnitude of a free flight time, i.e., the mean free
time $\tau_0$. We choose the latter approach, because an estimate of
the deviation from the exact CPR can be made at small $\gamma$, as a
{\it power series expansion} in $\gamma$. It is important to realize
at this point that as the density sets a time scale in the form of the
mean flight time $\tau_{0}$ between collisions, the actual
dimensionless small parameter corresponding to the shear rate is
$\tilde\gamma=\gamma\tau_{0}$.

We begin by constructing another matrix ${\mathbf H}_0(\Delta t)$ by
setting $\gamma=0$ but $\alpha_0\neq0$ in the explicit form of
${\mathbf H}(\Delta t)$ in Eqs.~(\ref{e9}-\ref{e4.3}), i.e.,
\begin{eqnarray*}
{\mathbf H}_0(\Delta t)\,=\,{\mathbf H}(\Delta
t)|_{\alpha_0\neq0,\,\gamma\,=\,0}\,.
\end{eqnarray*}
It is easy to show that ${\mathbf H}_0(\Delta t)$ satisfies the
equation
\begin{eqnarray*}
[{\mathbf H}_0(\Delta t)]^{\mbox{\scriptsize T}}\,{\mathbf
J}\,{\mathbf H}_0(\Delta t)\,=\,e^{-\alpha_{0}\Delta t}\,{\mathbf J}\,.
\end{eqnarray*}
Following Eq.~(\ref{e15}), we then form the matrix ${\mathbf L}_0(t)$
as
\begin{eqnarray}
   {\mathbf L}_0(t)={\mathbf H}_0\,(\Delta t_{s}){\mathbf M}_{i_sj_s}
	{\mathbf H}_0(\Delta t_{s-1})  \cdots {\mathbf
	M}_{i_1j_1}\,{\mathbf H}_0(\Delta t_{0})\,,
\label{e23}
\end{eqnarray}
such that {\it all\/} the ${\mathbf M}_{ij}$ matrices in
Eqs.~(\ref{e15}) and (\ref{e23}) are the same. Since both the
${\mathbf M}_{ij}$ and the ${\mathbf H}_0(\Delta t)$ matrices are now
$\mu$-symplectic with ${\mathbf J}$, so is ${\mathbf L}_0(t)$. As a
consequence, the logarithms of the eigenvalues of
${\mathbf\tilde{L}}_0(t)\,=\,[{\mathbf L}_{0}(t)]^{\mbox{\scriptsize
T}}{\mathbf L}_{0}(t)$ pair exactly.  This implies that if we arrange
the corresponding Lyapunov spectrum
\begin{eqnarray*}
 {\mathbf \Lambda}_0 &=& \lim_{t\rightarrow\infty}\left[
        {\mathbf\tilde{L}}_0(t)\right]^{1/(2t)}\,,
\end{eqnarray*}
in the decreasing order of magnitude as
$\lambda^{(0)}_{1}\geq\lambda^{(0)}_{2}\geq\ldots\geq\lambda^{(0)}_{6N}$,
then $\lambda^{(0)}_{i}+\lambda^{(0)}_{6N-i+1}=-\alpha_{0}$.

It is a simple exercise to show that ${\mathbf H}(\Delta t)-{\mathbf
H}_0(\Delta t)=O(\tilde\gamma^3)$, from which we conclude that for
$\Delta t=\tau=O(\tau_0)$
\begin{eqnarray}
	{\mathbf L}(\tau) = {\mathbf L}_0(\tau)\,[{\mathbf
	I}\,+\,\tilde\gamma^3 {\mathbf B}\,]\,,
\label{e26}
\end{eqnarray}
where the matrix ${\mathbf B}$ is of order 1 in $\tilde\gamma$ and
order 1 in $N$. Note that $\mathbf B$ contains higher powers of
$\tilde\gamma$ as well. Because it involves the matrix $\mathbf c$ and
contributions from collisions between spheres, ${\mathbf B}$ is not
proportional to $\mathbf I$, and hence, we cannot regard it simply as
a scalar factor (in which case the exact conjugate pairing would be
easy to obtain again).  Equation (\ref{e26}) implies that for
$\Delta{\mathbf\tilde L}(\tau)\equiv{\mathbf \tilde
L}(\tau)\,-\,{\mathbf \tilde L}_0(\tau)$
\begin{eqnarray}
   \Delta{\mathbf\tilde L}(\tau) &=& \tilde\gamma^3\,[\,{\mathbf
B}^{\mbox{\scriptsize T}}{\mathbf \tilde L}_0(\tau)\,+\,{\mathbf
\tilde L}_0(\tau){\mathbf B}\,] \,+\,\tilde\gamma^6{\mathbf
B}^{\mbox{\scriptsize T}}{\mathbf \tilde L}_0(\tau){\mathbf B}.
\label{e27}
\end{eqnarray}
{}From Eqs.~(\ref{e26}) and (\ref{e27}), we can now see that the
differences between ${\mathbf L}(\tau)$ and ${\mathbf L}_0(\tau)$, and
between ${\mathbf \tilde L}(\tau)$ and ${\mathbf \tilde L}_0(\tau)$
are small, by a relative order $\tilde\gamma^3$. Therefore the
logarithm of the eigenvalues of ${\mathbf L}(\tau)$ and ${\mathbf
L}_0(\tau)$ also differ by a term of order $\tilde\gamma^3$ in an
absolute sense.  If we now divide the logarithms of these eigenvalues
by the time $\tau$, we see that the finite time (for time $\tau$)
Lyapunov exponents, calculated from ${\mathbf\tilde L}_0(\tau)$ and
from ${\mathbf \tilde L}(\tau)$ (which we denote as
$\lambda^{(0)}_{i}(\tau)$ and $\lambda_{i}(\tau)$ respectively, for
$i=1,2\ldots6N$), differ by a term
$O(\tilde\gamma^3/\tau)=O(\gamma\tilde\gamma^2)$.

We make one further observation at this stage. The Lyapunov exponents
(even the finite time ones) are invariant under
$\gamma\rightarrow-\gamma$, so in a power series expansion
in $\tilde\gamma$ \cite{analytic}, the odd powers vanish. Hence, we
conclude that the logarithm of the eigenvalues of ${\mathbf L}(\tau)$
and ${\mathbf L}_0(\tau)$ must differ by a term of order
$\tilde\gamma^4$, i.e., the conjugate pairing of $\lambda_{i}(\tau)$'s
must be valid up to corrections of the form $\gamma\tilde\gamma^3$.

To explicitly extend this formalism to large $t$ and thereby obtain a
relation between $\lambda_{i}$s and $\lambda^{(0)}_{i}$s, we need to
sequentially concatenate a lot of ${\mathbf L}(\tau)$'s. In general,
these matrices neither commute with each other, nor with the ${\mathbf
B}$'s, which prevents us from explicitly demonstrating how the
deviation $[{\mathbf L}(t)-{\mathbf L}_0(t)]$ is built up. However, we
can argue in the following manner: ${\mathbf\tilde L}(t)$ and
${\mathbf\tilde L}_{0}(t)$ are positive definite and symmetric. This
allows us to express them in the form ${\mathbf\tilde
L}_{0}(t)=\exp({\mathbf A}_{0})$ and ${\mathbf\tilde
L}(t)=\exp({\mathbf A})$, where for large $t$, both the eigenvalues of
${\mathbf A}_{0}$ and ${\mathbf A}$ must behave $\sim t$.  From this
perspective, the difference between the Lyapunov exponents for
${\mathbf\tilde L}(t)$ and ${\mathbf\tilde L}_{0}(t)$ is related to
$({\mathbf A}-{\mathbf A}_{0})$. Since the difference between
${\mathbf\tilde L}(t)$ and ${\mathbf\tilde L}_{0}(t)$ has an explicit
prefactor of $\tilde\gamma^{3}$, so does ${\mathbf A}-{\mathbf
A}_{0}$.  Using the symmetry argument that the Lyapunov exponents have
to be even functions of $\gamma$, we obtain
\begin{eqnarray}
\lambda_i +  \lambda_{6N-i+1}\,=\, -\alpha_0 +
O(\gamma\tilde\gamma^3)\,.\quad i=1,\ldots6N
\label{e28}
\end{eqnarray}
For the largest and the most negative Lyapunov exponents, it has been
possible to show that they pair to $-\alpha_0$ plus corrections of
$O(\gamma\tilde\gamma^3)$ by means of a kinetic theory approach
\cite{vanZon_thesis,vZvB_unpublished}, based on the independence of
subsequent collisions of a sphere.  Likewise, one expects that in
subsequent time-intervals of $O(\tau_0)$, the ${\mathbf L}(\tau)$
matrices are not qualitatively much different from each other.
Therefore, we expect that the coefficient of the
$O(\gamma\tilde\gamma^3)$ term in Eq.~(\ref{e28}), to be of the same
order as that for a flight time $\tau=O(\tau_{0})$ [i.e. of the order
of ${\mathbf B}=O(1)$], and therefore Eq.~(\ref{e28}) to hold.

In summary, for the SLLOD equations with a constant $\alpha_0$
thermostat, the finite time Lyapunov exponents obey the CPR up to
$O(\gamma\tilde\gamma^3)$ when that time is of the order of the mean
flight time, and this is expected to hold for the infinite time
Lyapunov exponents too. Moreover, the iso-kinetic Gaussian thermostat
is equivalent to the constant multiplier thermostat in the
thermodynamic limit\cite{Zon_pre_99}, and hence one expects that with
an iso-kinetic Gaussian thermostat between collisions, the Lyapunov
exponent spectrum also exhibits $O(\gamma\tilde\gamma^3)$ deviations
from the CPR, in the thermodynamic limit. Finally, given that the
source of the CPR violation is basically the $\alpha_0\gamma{\mathbf
C}\vec{R}$ term in Eq. (\ref{e6}), one can argue that when the
gas particles interact with each other by means of a short-ranged,
repulsive potential with a constant multiplier thermostat, the 
violation of the CPR would also be at least of $O(\tilde\gamma^4)$ 
(for gas particles interacting with each other by means of a 
short-ranged, repulsive potential with an isokinetic Gaussian 
thermostat, the same results are expected) \cite{PZ_long}.

It is a pleasure to thank Prof. J. R. Dorfman and Prof. H. van Beijeren
for many useful and motivating discussions regarding this subject. D.\ P.
was supported by the research grants from ``Fundamenteel Onderzoek der
Materie (FOM)'' and that of Prof. J.\ R.\ Dorfman, NSF-PHY-9600428. R. v.\
Z. was supported by the research grant of Prof. H. van Beijeren, by FOM,
SMC and NWO Priority Program Non-Linear Systems, and by a grant from the
Natural Sciences and Engineering Research Council of Canada.

\ecols

\begin{thebibliography}{00}
\bibitem{ignorant} With CPR, the viscosity can be computed from two
pairing Lyapunov exponents, but numerically it is not very efficient.
\bibitem{otherG} There may exist other forms of ${\mathbf g}$
satisfying Eq.~(\ref{e10}), but Eq.~(\ref{e11}) is the simplest one
for which
${\mathbf G}^{2}\propto {\mathbf I}$.
\end{thebibliography}
\end{document}